\documentstyle[seceq,epsf]{ptptex}
\title{Axionic Extensions of the Standard Model}

\author{Jihn E. {\sc Kim}
 \footnote{E-mail address: jekim@phyp.snu.ac.kr}}

\inst{Physics Department and Center for Theoretical Physics, 
Seoul National University, Seoul 151-742, Korea}

\recdate{%
\today
}

\abst{After a brief review of the axion solution of the strong
CP problem, its supersymmetric extension and the superstring
axion are discussed. I also present one interesting cosmological
scenario, the axino-gravitino cosmology,
for the large scale structure of the universe.}

\begin{document}

\maketitle

\section{The Axion Solution of the Strong CP Problem}
The standard model describes the low energy phenomena very
successfully with 19 free parameters given by
\vskip 0.1cm
\begin{eqnarray*}
&{\rm Gauge\ sector }\ :\ \alpha_3,\ \alpha_2,\ \alpha_1,\ M_W,\ M_Z\\
&{\rm Higgs\ potential}\ :\ \lambda,\ M_H^2\\
&{\rm Leptons}\ :\ m_e,\ m_\mu,\ m_\tau\\
&{\rm Quarks}\ :\ {\rm 6\ masses\ +\ 4\ angles}\\
&{\rm Vacuum\ angle}\ :\ \theta_{QCD}
\end{eqnarray*}
\vskip 0.2cm
\noindent with two constraints, $M_H^2=\lambda(2/\pi)(M_W^2/\alpha_2), 
M_W^2=M_Z^2\cos^2\theta_W$.  It is known that the fundamental
problem of the standard model is to understand the origin of these
parameters. In fact, the progresses of particle physics during the
last two decades are along this line as shown below:
\vskip 0.2cm
\begin{tabular}{cc}
{\bf Parameters}& {\bf Extensions}\\
{\rm gauge couplings}& {\rm GUTs} \cite{rf:gut}\\
$\theta_{QCD}$& {\rm axions} \cite{rf:axion}\\
{\rm Higgs boson mass}& {\rm technicolor,\cite{rf:tech} SUSY 
\cite{rf:susy}} \\
{\rm fermion masses, etc.}&  $U(1)_{\rm
horizontal}$,\cite{rf:u1} etc.
\end{tabular}
\vskip 0.4cm

In this talk, I will concentrate on the parameter problem on $\theta_{QCD}$,
the so-called {\it the strong CP problem}.\cite{rf:rev} 

The strong interactions are described by quantum chromo dynamics (QCD),
in which quarks and gluons interact based on the $SU(3)_c$
gauge symmetry.  From gluon fields, we expect the following
gauge invariant terms
\begin{equation}
-{1\over 4g^2}F_{\mu\nu}^aF^{a\mu\nu}+{\bar\theta\over 32\pi^2}
F_{\mu\nu}^a\tilde F^{a\mu\nu}
\end{equation} 
where $\tilde F_{\mu\nu}^a={1\over 2}\epsilon_{\mu\nu\rho\sigma}
F^{a\rho\sigma}$.
If there is no massless quark, the $\bar\theta$ term is
present and physical.  However,
a massless quark makes the $\bar\theta$ term unphysical.
A physical $\bar\theta$ is given by
\begin{equation}
\bar\theta\ =\ \theta_{QCD}+\theta_{QFD}
\end{equation}
with the self-explanatory suffices.

As a simple example, consider the $F\tilde F$ term in 
a $U(1)$ gauge theory,
\begin{equation}
{1\over 2}\epsilon^{\mu\nu\rho\sigma}(\partial_\mu A_\nu-\partial_\nu
A_\mu)(\partial_\rho A_\sigma-\partial_\sigma A_\rho)
=2\partial_\mu(\epsilon^{\mu\nu\rho\sigma}A_\nu \partial_\rho A_\sigma)
\end{equation}
which is a total divergence, but can contribute to the
action if the surface term at spatial infinity is nonvanishing. 
Because the $U(1)$ gauge field does not
have nonlinear interactions, configurations of the $U(1)$ gauge field
does not introduce a nonvanishing contribution at spatial infinity. 
Thus the $F\tilde F$ terms in $U(1)$ gauge theories can be neglected.

In nonabelian gauge theories, the $F\tilde F$ term is also a
total divergence.
But in nonabelian gauge theories, $A_\mu$ at infinity can be important
due to the instanton solution (through the nonlinear term).\cite{rf:inst}  
Namely, one cannot neglect $\theta F\tilde F$ term in QCD.
The CP property of the $F\tilde F\ (\sim {\bf E\cdot B})$ is
different from that of $F^2\ (\sim {\bf E^2-B^2})$, since $\bf E\rightarrow
-E$ and $\bf B\rightarrow B$ under CP transformation.  Thus $\theta
F\tilde F$ term violates the CP invariance.  Since it occurs through the
strongly interacting gluon fields, the corresponding coupling $\bar\theta$
must be tuned to a very small value so that it is consistent with
the upper bound of the neutron electric dipole moment,\cite{rf:edm}
\begin{equation}
|\bar\theta| < 10^{-9}.
\end{equation}
 
\noindent
{\bf Why is $\bar\theta$ so small?}

The $\bar\theta$ parameter consists of two parts,
$\bar\theta=\theta_{QCD}+\theta_{QFD}$ where $\theta_{QFD}$ is the
contribution when electroweak CP violation is taken into account,
and is expressed as $Arg. Det. M_q$.
The $\theta_{QCD}$ is the original parameter descended from high energy
scale before taking into account the electroweak symmetry breaking. 
The smallness of $\bar\theta$ can results from either\\
\vskip 0.1cm
\indent i) Fine tune $\bar\theta=0$, which is not a understanding
since it relies on the miraculous cancellation of
$\theta_{QCD}$ and $\theta_{QFD}$, or\\
\indent ii) Natural solutions in which
one insists that (a) the Lagrangian is CP invariant, (b) CP violation
in weak interactions are generated by the spontaneous symmetry
breaking mechanism, and (c) loop corrections to $\theta_{QFD}$
are negligible, or\\
\indent iii) Dynamical solutions, which is the axion solution 
\cite{rf:axion} and can be important in cosmology,\cite{rf:axcos} or\\
\indent iv) Massless $u$-quark possibility.\cite{rf:uq}\\
\vskip 0.1cm
Both iii) and iv) base their arguments on global symmetries, which
is difficult to realize in string models.
\vskip 0.5cm

\noindent
{\bf Axion solves the strong CP problem.}

Below the $SU(2)\times U(1)$ symmetry breaking scale, the
QCD Lagrangian can be represented as
\begin{equation}
{\cal L}\ =\ -{1\over 4}F_{\mu\nu}^aF^{a\mu\nu}
+\bar q(i \slash{\hskip-0.27cm D}-M_q)q +\bar\theta \{F\tilde F\}
\end{equation}
where $M_q$ is real, diagonal, and $\gamma_5$-free matrix, and
$\{F\tilde F\ \}\equiv (g^2/32\pi^2)F^a_{\mu\nu}\tilde F^{a\mu\nu}$.
Now we understand that $\bar\theta$ is a dynamical field in axion physics,
but for a moment let us treat it as a parameter.  A simple proof
that $\bar\theta=0$ is the minimum of the potential is given by
Vafa and Witten.\cite{rf:vafa}  Expressing the generating functional
after integrating out the quark fields, we obtain 
the partition function in the Euclidian space
\begin{equation}
\int [dA_\mu] \prod_i Det. (\slash{\hskip-0.27cm D}
+m_i)\exp\left(-\int d^4x[{1\over 4g^2}F^2
-i\bar\theta\{F\tilde F\}]\right).
\end{equation}
Note that the $\bar\theta$ term is pure imaginary.
The Euclidian Dirac operator satisfies
$i\slash{\hskip-0.27cm D}\psi=\lambda\psi$ and $i
\slash{\hskip-0.27cm D}(\gamma_5\psi)=-\lambda(\gamma_5\psi)$.
Therefore, if $\lambda$ is a real eigenvalue of 
$i\slash{\hskip-0.27cm D}$, so is $-\lambda$,
implying,
\begin{equation}
Det. (\slash{\hskip-0.27cm D}+m_i)= \prod_{\lambda}(-i\lambda+m_i)
=m_i^{N_0}\prod_{\lambda>0}(m_i^2+\lambda^2)>0
\end{equation}
where $N_0$ is the number of zero modes.  Therefore, using the
Schwarz inequality we obtain
\begin{eqnarray*}
e^{-\int d^4xV[\bar\theta]}\ &\equiv\ \left|\int [dA_\mu]\prod_i Det.
 (\slash{\hskip-0.27cm D}+m_i)
e^{-\int d^4x[{1\over 4g^2}F^2-i\bar\theta\{F\tilde F\}]}\right|\\ 
&\le\ \int [dA_\mu]\left|\prod_i Det. 
(\slash{\hskip-0.27cm D}+m_i)e^{-\int d^4x[{1\over 4g^2}F^2-i
\bar\theta{F\tilde F}]}\right|\\
&=\ \int [dA_\mu]\prod_i Det. 
(\slash{\hskip-0.27cm D}+m_i)e^{-\int d^4x {1\over 4g^2}F^2}\\
&=\ e^{-\int d^4x V[0]},
\end{eqnarray*}
which implies
\begin{equation}
V[\bar\theta]\ge V[0].
\end{equation}
A schematic behavior of $V[\bar\theta]$ is shown below.
\begin{figure}
\epsfxsize= 12 cm
\epsfysize= 7 cm
\centerline{\epsffile{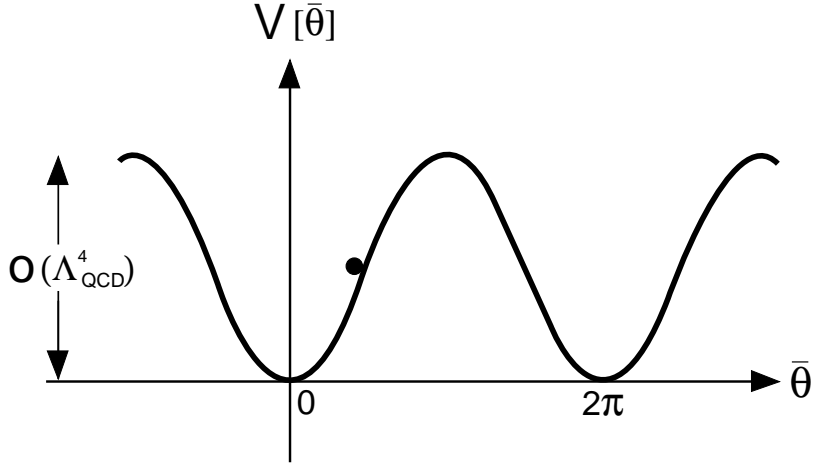}}
\caption{$V[\bar\theta]$ versus $\bar\theta$.}
\end{figure}
\vskip 0.5cm

The maximum points of Fig. 1 gives $(2Z/(1+Z)^2)f_\pi^2m_\pi^2$ from
the current algebra estimate,\cite{rf:current} 
where $Z=m_u/m_d$ in terms of the
current quark masses.  Because $Z\sim \int [dA_\mu](\cdots)
e^{i\bar\theta\int d^4x\{F\tilde F\}}$ and $\int d^4x \{ F\tilde
F\}={\rm integer}$, $\bar\theta$ is a periodic variable with
periodicity of $2\pi$.  Even though $V[\bar\theta]$ takes the above
form, any $\bar\theta$ as shown with a dot in Fig. 1 
will be allowed as any magnitude for the
fine structure constant $\alpha_{em}$ is allowed theoretically. Thus,
for a nonzero $\bar\theta$, CP invariance is violated in strong
interactions.  The axion solution for the strong CP problem
is to identify $\bar\theta$ as a dynamical field $a$ instead of a
coupling constant,
\begin{equation}
\bar\theta\ =\ {a\over F_a}.
\end{equation}
If $\bar\theta$ is a coupling, different $\bar\theta$'s describe
different worlds (or theories), but in the axion world different
$\bar\theta$'s mean different vacua in the same theory.  The
vacuum is corresponding to $\bar\theta=0$ as depicted in Fig. 1.

An important feature to be satisfied in the axion models is that
{\it the axion $a$ does NOT have any potential except that coming
from $\bar\theta\{F\tilde F\}$, or the mechanism does not work.}

To make $\bar\theta$ dynamical, one must have a mechanism to 
introduce a scale $F_a$:\vskip 0.3cm
\indent (i) $a$ is the {\it Goldstone boson} of a spontaneously
broken $U(1)_A$ symmetry where the $U(1)_A$ must have the QCD 
anomaly $\partial_\mu j^\mu\sim F\tilde F$ so that $(a/F_a)F\tilde F$
coupling arises,\cite{rf:axion} or\\
\indent (ii) $a$ is a fundamental field in string models, and the
scale $F_a$ arises from compactification.  From low energy point
of view the nonrenormalizable interaction term is present.  The
model-independent axion in string models belongs to this
category,\cite{rf:staxion} or\\
\indent (iii) $a$ is a composite field, and $F_a$ arises at the
confining scale.\cite{rf:compaxion}
\vskip 0.5cm

\noindent {\bf Axion potential and cosmology}

The invisible axion, due to its small couplings, has a dramatic
cosmological consequence.\cite{rf:axcos}  (A similar argument holds for the
Polonyi problem.\cite{rf:pol})
\begin{figure}
\epsfxsize= 15 cm
\epsfysize= 3 cm
\centerline{\epsffile{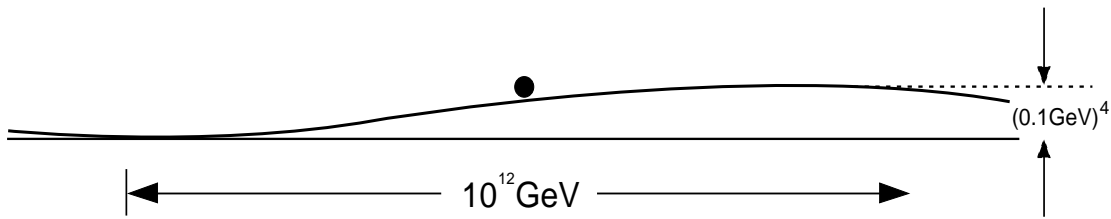}}
\caption{The almost flat axion potential.}
\end{figure}
\vskip 0.5cm
In Fig. 2, we have shown the axion potential schematically.  Actually,
it is extremely flat as emphasized in Fig. 2 
when the temperature of the universe $T$ is
greater than 1 GeV.  Only when the expansion rate of the universe
is sufficiently slowed down at $H\simeq m_a$, the axion potential
is felt and the vacuum starts to oscillate around $\bar\theta=0$.
Since the axion lifetime is greater than $10^{20}$ times the
age of the universe, the collective oscillation mode does not
die out by the axion decay.  Its amplitude of oscillation is shrunk only
by the expansion of the universe, and corresponds to $\bar\theta\sim
10^{-20}$ now which is safely within the experimental upper bound.

For $V[a]\simeq (1/2)m_a^2a^2$, the energy density of axion
$\rho=(1/2)\langle\dot a\rangle ^2+V$ behaves like that of nonrelativistic
particles.  This is the reason that the axion is classified as the
cold dark matter candidate, but it is only true for the classical
motion.  The cold axion energy density satisfies
\begin{equation}
{d\over dt}\rho\ =\ -3H (\rho+p)
\end{equation}
where $p=(1/2)\langle \dot a\rangle ^2-V$. Parametrizing $\langle
a\rangle$ as
\begin{equation}
\langle a(t)\rangle\ =\ A(t)\cos m_at,
\end{equation}
one obtains
\begin{equation}
{d\over dt}(m_aA^2)\ =\ -3H(m_aA^2) \ \ \rightarrow\ \
{(m_aA^2)_f\over (m_aA^2)_i}\ =\ \left({R_f\over R_i}\right)^{-3}.
\end{equation}
In the adiabatic expansion, the above equation shows that $m_aA^2$
is the number density, and hence the energy density becomes
\begin{equation}
\rho_a\ =\ m_an_a\ =\ {1\over 2}m_a^2A^2.
\end{equation}
The axion vacuum, denoted as a dot in Fig. 2 starts to roll down the
hill when the temperature of the universe is lowered to $\sim$ 1
GeV.  From then on the axion vacuum oscillate around the minimum
$\bar\theta=0$.  One can estimate the energy density of this
collective motion of cold axions which is
\begin{equation}
\Omega_a\ =\ 2.1\times 10^7 h^{-2}_{1/2}\left({{\rm GeV}\over T_1}\right)
\left(T_\gamma\over 2.72\ {\rm K}\right)^2\left({A(T_1)\over F_a}\right)
{F_a\over M_{Pl}}
\end{equation}
where $h_{1/2}$ is the Hubble parameter in units of 50 $\rm km\cdot
s^{-1}\cdot Mpc^{-1}$, $T_1\simeq 1$ GeV is the scale when axion vacuum
starts to roll, and $A(T_1)$ is the axion amplitude.  From $\Omega_a
\le 1$, we obtain
\begin{equation}
F_a\ \le\ 0.5\times 10^{12} h_{1/2}^2\ \rm GeV.
\end{equation}
We also have $|\bar\theta|_{\rm now}<10^{-20}$.

\section{Supersymmetric Extension}

Last 20 years in theoretical particle physics is dominated by the
study of
supersymmetry which has been suggested as the best candidate
toward the solution of the hierarchy problem.
Therefore, axions should be understood in this framework if
supersymmetry is the fundamental symmetry of nature.  The
starting point for phenomenology is the minimal supersymmetric
standard model (MSSM) where all known particles of the standard model
accompany their superpartners.  In addition, two Higgs doublets
and their superpartners are introduced : $H_1$ and $H_2$.  $H_1$
couples to charged leptons and $d$-type quarks and $H_2$ couples
to $u$-type quarks.  In MSSM, all dimensionless couplings come from
supersymmetric terms.  For dimensionful parameters, there are two
categories.  There exists soft supersymmetry breaking parameters
appearing in scalar mass terms and gaugino mass terms.  These soft
supersymmetry breaking parameters have magnitudes comparable
to the gravitino mass 
\begin{equation}
m_{3/2}\ \simeq\ {M_I^2\over M_{Pl}}.
\end{equation}
For supersymmetry to be responsible for the hierarchy problem,
$m_{3/2}\sim M_W$ and hence there must be an intermediate scale
$M_I\simeq 10^{10}$ GeV $\sim$ $10^{11}$ GeV.  Another dimensionful
parameter in MSSM appears in the superpotential as the
$\mu$ term 
\begin{equation}
W_\mu\ =\ \mu H_1H_2.
\end{equation}
This $\mu$ term is supersymmetric, and the intermediate scale
$M_I$ is not directly related to $\mu$.
Therefore, the parameter $\mu$ at Planck scale can be of order $M_{Pl}$.
Then, the electroweak scale $M_W$ cannot be generated.  On the
other hand, the magnitude of $\mu$ is 
known to be nonzero from high energy 
experiments.  In addition, if it were zero, one expects an
electroweak scale axion which has been ruled out 
phenomenologically.\cite{rf:rev}  Also, $\langle H_1\rangle=0$
results from $\mu=0$, which is in contradiction with
making $Q=-1/3$ quarks massive.
Therefore, $\mu$ is expected to be of order electroweak scale
$\sim m_{3/2}$.  How $\mu\sim M_W$ can occur is the so-called
{$\mu$ \it problem}.\cite{rf:mu}

If one introduces a Peccei-Quinn symmetry in supergravity, one
can generate a reasonable $\mu$ term.  For example, a nonrenormalizable
term 
\begin{equation}
W_\mu\ =\ {1\over M_{Pl}}S_1S_2H_1H_2
\end{equation}
can be considered.  Here $S_{1,2}$ are gauge singlets.
Because of the Peccei-Quinn symmetry, one assigns
charge 1 to $H_1$ and $H_2$ fields.  Then the sum of the $S_1$ and
$S_2$ charges must be -2.  Therefore, by giving nonzero vacuum
expectation values to $S_1$ and $S_2$ at $\sim 10^{12}$ GeV, one
breaks the Peccei-Quinn symmetry and generate a correct order for
the $\mu$ term.  In the process, there results an invisible axion.

A general K$\ddot {\rm a}$hler potential is restricted by the
reality condition only.\cite{rf:gm}  Therefore, one can write
\begin{equation}
{\cal G}\ =\ \sum_{a}y^ay^{\dagger a}+zz^\dagger+
(\lambda H_1H_2+\rm h.c.)=\cdots
\end{equation}
with observable sector fields $y^a$ and a hidden sector field $z$.
Of course, such $H_1H_2$ term can be present in the K$\ddot {\rm a}$hler
potential, but one has to explain why 
$M_{Pl}H_1H_2$ is not present in the
superpotential.  This can be done only by a symmetry 
argument.\cite{rf:musym}

\section{Superstring Axion}

The N=1 supergravity theory in D=10 needs a supergravity
multiplet
\begin{equation}
B_{MN}\ \ (M,N=0,1,\cdots,9);\ \ \ \ M, N\ \rm antisymmetric.
\end{equation}
The D=10 supergravity models obtained from superstring models
also needs the $B_{MN}$ field.  After compactifying 6 internal
dimensions, the field strength of $B_{\mu\nu}\ (\mu,\nu,\rho=
0,1,2,3)$ is sometimes duality transformed to see its physical
implications,
\begin{equation} 
H_{\mu\nu\rho}\ =\ 16\pi^2\epsilon_{\mu\nu\rho\sigma}J_1^\sigma
\end{equation}
where $J_1^\sigma$ is a gauge invariant d=4 axial vector current,
\begin{equation}
\partial_\mu J_1^\mu\ =\ {1\over 32\pi^2}\left(
{\rm Tr}\ R_{\mu\nu}\tilde R^{\mu\nu}-{1\over 30}
{\rm Tr}\ F_{\mu\nu}\tilde F^{\mu\nu}\right).
\end{equation}
We define the model-independent axion $a_{MI}$ as the
linear combination of $a_1$
\begin{equation}
F_1\partial^\mu a_1=J_1^\mu={1\over 96\pi^2}\epsilon^{\mu\nu\rho\sigma}
H_{\nu\rho\sigma}
\end{equation}
and $B_{mn}\equiv B_A$ obtained from the process of compactification,
\begin{eqnarray*}
{\cal L}_{eff}\ &=\ {1\over 2}(\partial_\mu a_1)^2
+{1\over 2}\sum_{A=1}^{N}(\partial_\mu B_A)^2-{a_1\over 32\pi^2F_1}
{1\over 30}F\tilde F\\
&-{1\over 32\pi^2F_2}(\sum_A \Gamma^AB_A)({1\over 30}
\sum_aC^a{\rm Tr}_aF\tilde F)
\end{eqnarray*}
where $\Gamma^A$ and $C^a$ depend on compactification.  Note that
$F_1$ can be estimated by comparing with Newton's constant
\cite{rf:choi}
\begin{equation}
F_1\ =\ {g^2\over 192\pi^{5/2}}M_{Pl}\ \simeq\ 1.5\times 10^{15} \ 
\rm GeV.
\end{equation}
Actually, due to the $F_2$ term the model-independent axion decay 
constant $F_a$ is not exactly $F_1$. $F_1$ gives an idea on the 
magnitude of
$F_a$.  For example, the $E_8\times E^\prime_8$ heterotic string
compactified with $b_{2,0}=b_{0,2}=0, b_{1,1}=1$ leads to the
following coupling \cite{rf:choik}
\begin{equation}
{1\over 32\pi^2F_1}a_1(F\tilde F+F^\prime\tilde F^\prime)
+{1\over 32\pi^2F_2}a_2(F\tilde F-F^\prime\tilde F^\prime)
\end{equation}
from which we obtain
\begin{equation}
F_a\ =\ {1\over 2}\sqrt{F_1^2+F_2^2}\ >\ 0.7\times 10^{15}\ \rm GeV.
\end{equation}
In general, the model-independent axion scale falls in the
compactification scale $\sim 10^{17}$ GeV.
Therefore, the scale of $a_{MI}$ is too large: the universe
collapses before life was born if there is no extra confining
gauge group, or the strong CP problem is not solved by $a_{MI}$ if
there exists an extra hidden sector confining gauge group.
Thus it is very important to find a solution of the strong CP
problem in string models.

Many compactification schemes introduce an anomalous $U(1)$ gauge
symmetry.\cite{rf:iq}  
This anomalous $U(1)$ has been used to break extra $U(1)$
gauge symmetries appearing in many 4D string models.
The anomaly is not troublesome in string models
due to the model-independent axion $a_{MI}$ 
which makes the $U(1)$ gauge boson massive by becoming its
longitudinal degree.  Thus we see that the old 't Hooft mechanism
is operative here,
\begin{eqnarray*}
&a_{MI}\ (nonlinearly\ realized\ global\ symmetry)+
anomalous\ U(1)\ gauge\ symmetry\\
&\rightarrow U(1)\ global\ symmetry
\end{eqnarray*}
Comparing the Lagrangians of spontaneously broken $U(1)$ gauge
theory
\begin{equation}
{\cal L}\ =\ D_\mu\phi^* D^\mu\phi\ =\ {1\over 2}
(\partial_\mu a)^2-evA_\mu \partial^\mu a
\end{equation}
and the anomalous $U(1)$ gauge theory with $a_{MI}$
\begin{equation}
{\cal L}\ =\ -{1\over 4}F^2+{1\over 2}(\partial_\mu a_{MI})^2
+mA_\mu \partial^\mu a_{MI}+\cdots,
\end{equation}
we obtain $M_A^2=m^2$.  Below the compactification
scale $m$, the gauge boson $A_\mu$ is decoupled and a global
$U(1)$ symmetry is present.  Therefore, {\it if there is no
extra confining gauge group, the axion decay constant can be
lowered due to this new global symmetry which acts as an
anomalous $U(1)$ Peccei-Quinn symmetry.}\cite{rf:mich}  
However, a popular scenario for supersymmetry breaking is to
introduce an extra confining gauge group at $\sim 10^{13}$
GeV scale, the quantum hidden sector dynamics (QHD).  In this
case, the global symmetry produced by the 't Hooft mechanism
is explicitly broken by the quantum effects of QHD and we lack the 
axion needed to settle $\bar\theta$ at zero.  To solve the
strong CP problem, one assumes an approximate global symmetry in this
case.\cite{rf:chun}

Before discussing the axino-gravitino cosmology, consider 
\begin{equation}
{\rm Standard\ Model}\otimes SU(N)_{\rm hidden\ sector}.
\end{equation}
The model-independent axion $a_{MI}$ has the following
potential
\begin{equation}
V[a_{MI}]\ =\ \Lambda^4_{QCD}(1-\cos\bar\theta)
+\epsilon\Lambda^4_h (1-\cos\theta^\prime)
\end{equation}
where $\bar\theta\equiv (a_{MI}/F_a)+\theta_{QFD}$ and
$\theta^\prime\equiv a_{MI}/F_a$. Because the QHD scale $\Lambda_h
\gg\Lambda_{QCD}$, $\bar\theta$ is not settled at 0.  But if
$\epsilon=0$, the $\Lambda^4_{QCD}$ term dominates and $\bar\theta=0$
is the minimum of the potential.  Therefore,if a sufficiently small
$\epsilon$ (of order $<10^{-9}$) can be found, and
then the model is allowed.  One obvious solution
toward negligible $\epsilon$ is to introduce a massless QHD
quark.  The difficulty of introducing a massless QHD quark is to
introduce another global symmetry, not related to the $a_{MI}$,
in string models.  Thus the strong CP problem in superstring
models does not have an obvious solution with $\epsilon=0$ yet.
But there exists a possibility of having
a very small $\epsilon$ from the consideration of supersymmetry
and a global $U(1)$ resulting from 
the anomalous gauge $U(1)$.\cite{rf:eps}

\section{Axino-gravitino Cosmology}

Axion models in supergravity theories necessarily introduce
the gravitino $\tilde G$ and the axino $\tilde a$ whose masses
we denote as $m_{3/2}$ and $m_{\tilde a}$, respectively.  Of course,
the scalar partner of $\tilde a$ is present and it is sometimes
called saxion $s$.\cite{rf:rtw}\footnote{In the literature, saxino
(scalar partner of axino) has been used also.\cite{rf:saxino}}  
The saxion mass is of order $m_{3/2}$.  Before
discussing the axino-gravitino cosmology, let us briefly comment
some aspects of the gravitino mass effect on cosmology.

For $m_{3/2}\ge 1$ TeV, this region of the gravitino mass is
obtained from the $^4$He abundance calculation so that
the nucleosynthesis must starts from the scratch after the
reheating by gravitino decay.\cite{rf:sw}   The gravitino lifetime is
extremely long,
\begin{equation} 
\tau (m_{\tilde \gamma, gluino}\ll  m_{3/2})\ \simeq\
4.4\times 10^7\ {\rm s}\ \left({100\ {\rm GeV}\over m_{3/2}}\right)^3
\end{equation}
which is the reason that the gravitino is important in cosmology.

For 20 GeV $<\ m_{3/2}\ <$ 1 TeV, the gravitino decayed sometime
between the epochs of nucleosynthesis and recombination.  Even if one 
inflates away the primodial gravitinos, $\tilde G$ would have been
produced thermally.  If these thermally produced gravitinos were
too abundant, they would have destroyed the precious deuterium. 
This consideration restricts that the reheating temperature after
gravitino decay must be bounded\cite{rf:ekn}
\begin{equation}
T_{R,max}\ \simeq\ 10^{11}\sqrt{{m_{3/2}\over 100 \ \rm GeV}}
\ {\rm GeV}.
\end{equation}
The above bound is obtained by considering the scattering
cross section at zero temperature.  In cosmology, however,
high temperature effect may be important, since plasma of
particles does not respect supersymmetry.  But a
definite conclusion on this topic is premature.\cite{rf:fis}

In addition, saxion $s$ affects the evolution of the universe by
injecting more particles when it decays.  In this case, the
bound on the axion decay constant $F_a$ can be raised a 
bit.\cite{rf:saxino}
\vskip 0.5cm

\noindent {\bf Gravitino mass in the eV range}

Finally, let us discuss the axino-gravitino cosmology.
So far, we discussed supergravity models with a hidden sector
confining force, $\Lambda_h\sim 10^{13}$ GeV, which is the
source of supersymmetry breaking.  However, if supersymmetry is
broken by supercolor at $\Lambda_{supercolor}\simeq 1 \sim 10^3$ 
TeV,\cite{rf:dn}
i.e. with $m_{3/2}\simeq 0.4\times 10^{-3}\ {\rm eV}\ \sim\ 0.4\
{\rm keV}$, then axino $\tilde a$ can decay to gravitino $\tilde G$
plus axion,   
\begin{equation}
\tilde a\rightarrow \tilde G+a.
\end{equation}
This process might be very important in cosmology since the
lifetime of the axino is falling in the cosmologically interesting 
region.  Actually, this kind belongs to the cosmological scenario
with {\it late decaying particles}.  Cosmology with late
decaying particles was considered first by Bardeen, Bond and
Efstathiou in 1987,\cite{rf:bbe} and the now-dead 17 keV neutrino
was used to realize this scenario.  After the finding that
CDM models are in trouble with the COBE data of large scale
structure of the universe, Chun, Kim and Kim \cite{rf:ckk} first
considered the axino-gravitino cosmology, not knowing the old work of
Bardeen et al.\cite{rf:bbe}  
Recently, the late decaying particle cosmology is
studied by many groups primarily using one of the neutrinos as the
late decaying particle.\cite{rf:nu}

In the axino-gravitino cosmology, the axino lifetime is estimated to
be\cite{rf:saxino}
\begin{equation}
\tau_{\tilde a}\ =\ {96\pi M^2 m_{3/2}^2\over m_{\tilde a}^5}\ 
=\ 1.2\times 10^{12}\ {\rm s}\ \left({{\rm MeV}\over M_{\tilde a}}
\right)^5\left({m_{3/2}\over \rm eV}\right)^2
\end{equation}
from the Lagrangian ${\cal L}=(1/M)\bar
\psi_\mu\gamma^\nu\partial_\nu z^*\gamma^\mu\tilde a$ 
where $M=M_{Pl}/8\pi$.
The axino decoupling temperature is calculated mainly from the
process $q+\bar q\leftrightarrow \tilde a+\tilde g$,\cite{rf:rtw}
\begin{equation}
T_{\tilde a D}\ \simeq\ 10^{11}\ {\rm GeV}\ \left({F_a\over 10^{12}\
{\rm GeV}}\right)^2\left({0.1\over \alpha_c}\right)^3.
\end{equation}
\vskip 0.5cm
\begin{figure}
\epsfxsize= 12 cm
\epsfysize= 13 cm
\centerline{\epsffile{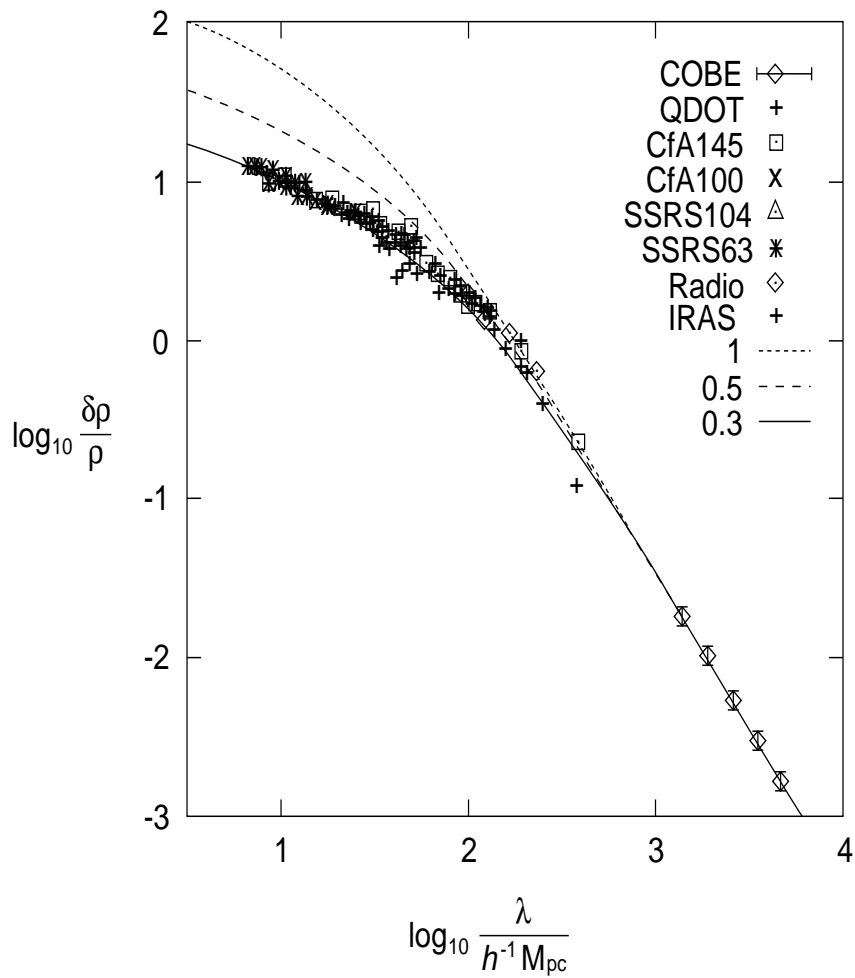}}
\caption{log$_{10}(\delta\rho/\rho)$ versus log$_{10}(\lambda/h^{-1}
{\rm Mpc})$.  The solid line corresponds to the $\Omega h=0.3$ CDM 
model, while the dashed and dotted lines corresponds to the
$\Omega h=0.5$ and 1 CDM models, respectively.}
\end{figure}

\noindent {\bf Dark matter and structure formation}

We envision that cold axions remain as dark matter, and gravitinos
and hot axions produced by the axino decay remain as hot dark matter.
Therefore, this late decaying particles with cold axions can mimick
the mixed dark matter model.\cite{rf:mixed} 
The cold dark matter (CDM) model was
successful in 80's.  But the normalization determined by the COBE
data necessiated modifications of the CDM models.   The study of
large scale structure is greatly simplified if we use the evolved
fluctuation spectrum calculated by Davis et al,\cite{rf:davis}
\begin{equation}
|\delta^2_k|^2\ =\ {Ak\over (1+\alpha k+\beta k^{3/2}+\gamma
k^2)^2}
\end{equation}
where $\alpha=1.7\ell,\beta=9.0\ell, \gamma=1.0\ell$ with
\begin{equation}
\ell\ =\ (\Omega h^2)^{-1}\theta^{1/2}\rm Mpc.
\end{equation}
Here, $\theta$ is determined by relativistic particles and
the old CDM model corresponds to $\theta=1$. In Fig. 3, we
present several CDM models with different values of $\Omega h$.

\begin{figure}
\epsfxsize= 12 cm
\epsfysize= 10 cm
\centerline{\epsffile{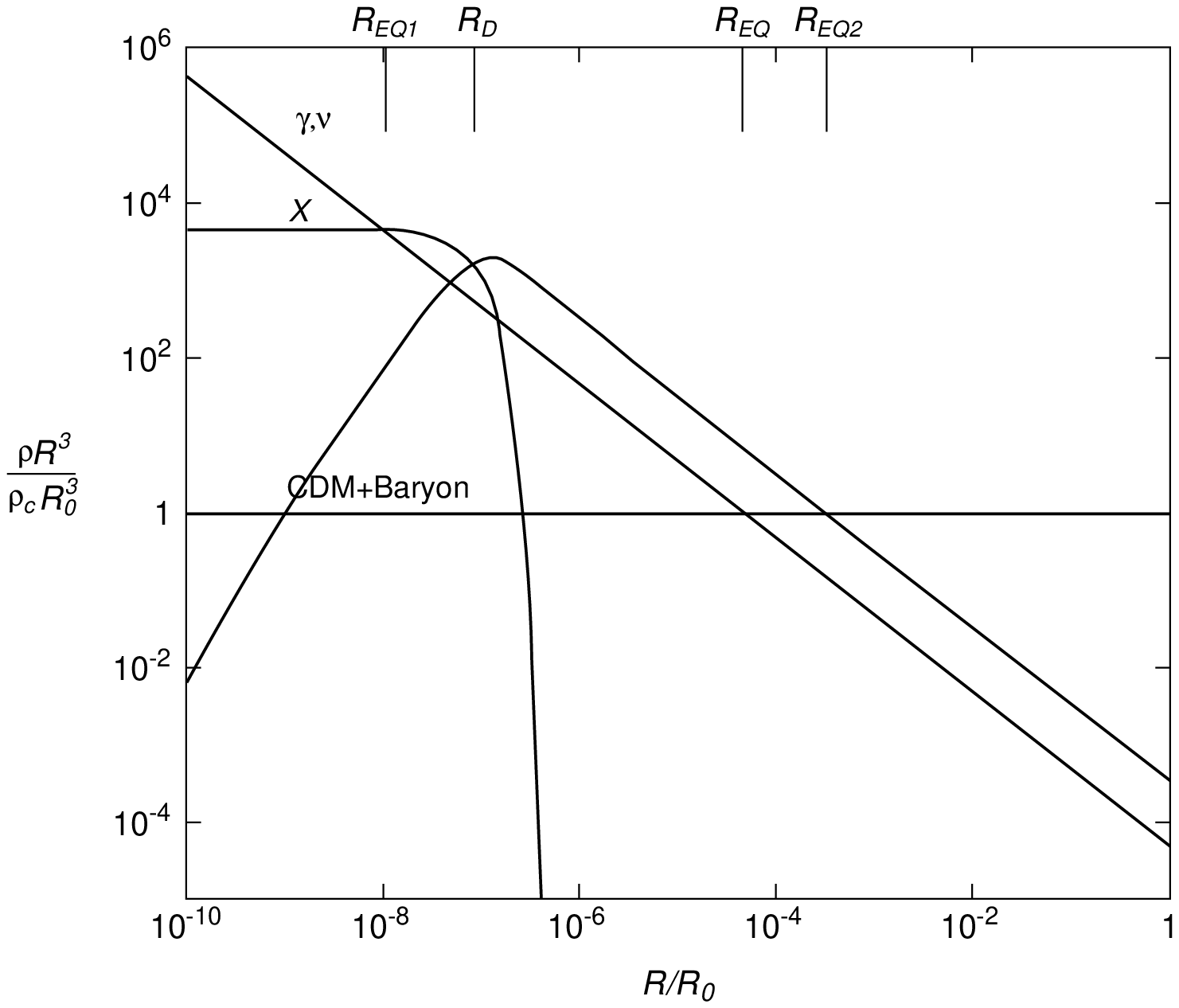}}
\caption{Scale factor versus energy densities of radiation
$(\gamma,\nu)$, axino ($\tilde a$), decay products of
axino ($\tilde G, a$) and cold axion ($\langle a
\rangle$) for $m_{\tilde a}=10$ MeV, $m_{3/2}=76$ meV
and $\rho_{\langle a\rangle 0}=\rho_c$.}
\end{figure}

Successful fits corresponds to
\vskip 0.5cm
\indent (i) $\Omega_{\rm cosmological\ constant}\simeq 0.8,\
\Omega_{CDM}\simeq 0.2$, or\\
\indent (ii) $\Omega_{CDM}\simeq 0.7,\ \Omega_{HDM}\simeq 0.3$, or\\
\indent (iii) $\Omega_{CDM}= 0.2\sim 0.3$, or\\
\indent (iv) Large biasing or antibiasing, or\\
\indent (v) Initial fluctuation spectrum with less power at small scales.
\vskip 0.5cm

Note that the effect of Case (iii) can be mimicked if we build a
model with $\theta\ne 1$ but with $\Omega=1$.  It is in effect
a CDM plus HDM model since $\theta\ne 1$ needs hot dark matter
except photons and neutrinos.  Instead of $\ell$, let us define
\begin{equation}
\lambda_{EQ}\ \simeq\ 30(\Omega h^2)^{-1}\theta^{1/2}\rm Mpc
\end{equation}
where
\begin{equation}
\theta\ =\ {\rho({\rm relativistic\ particles})\over 1.68\rho_\gamma}.
\end{equation}
Here, the denominator corresponds to neutrinos plus photons and the
numerator corresponds to all hot matters at present.  
A larger $\theta$ or a larger $\lambda_{EQ}$ with $\Omega=1$, 
which is required in most inflationary models,
can give the same $\ell$ given by
Case (iii).  Case (iii) with $\Omega\ne 1$, 
even though it gives a correct large scale
structure, is not welcome in inflationary models.  Thus a late decaying
particle scenario mimicking Case (iii) with $\Omega h_{1/2}=1$ and 
$\theta=1.8$ is more welcome in inflationary 
models.  In Fig. 4 we present $\rho$ versus $R$ (the scale factor)
plot.  From this figure, we note that for $R>R_{EQ2}$ the universe is
matter dominated and galaxies are formed.  However, we also note that
between $R_{EQ1}$ and $R_D$ (corresponding to the time of axino decay)
there is another epoch of matter domination.  The large scale structures
corresponding to this epoch is the size of globular clusters.  It will
be interesting if this turns out to be true.

In Fig. 5 we present the allowed regions of $m_{3/2}$ and $m_{\tilde a}$,
obtained from various cosmological data.  The axino-gravitino 
cosmology giving a desirable large scale structures discussed
above is denoted as a heavy dotted line.

\begin{figure}
\epsfxsize= 15 cm
\epsfysize= 11 cm
\centerline{\epsffile{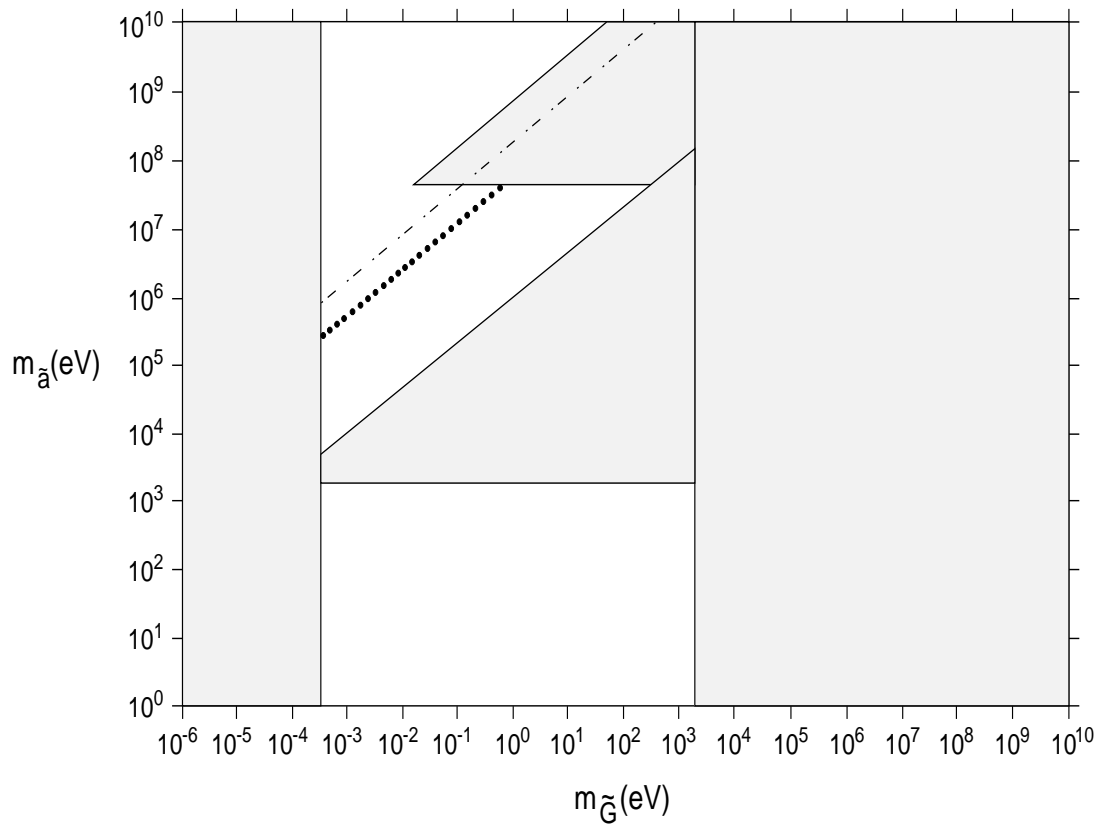}}
\caption{Allowed region in axino mass--gravitino mass.  Grey regions
are excluded by the closure bound and the nucleosynthesis bound.  Successful
fit to COBE data corresponds to the heavy dotted line.}
\end{figure}

\section{Conclusion}

The $\theta_{QCD}$ parameter problem of the standard model leads to
an axionic extension of the standard model.  Thus, one can consider
the invisible axion $a$ in addition to the standard model particles.
From the data on SN1987A and cosmological energy density we have an
axion widow $10^{10}\sim 10^{12}$ GeV.  It is possible to
introduce this range of $F_a$ in gauge
models without supersymmetry.  In supersymmetric models, one has
the $\mu$ term which can be introduced by a spontaneously broken
Peccei-Quinn symmetry.  In superstring models, the invisible axion
satisfying the experimental bound is more difficult to introduce.
Without an extra confining force,
the model-independent axion $a_{MI}$ can be made as the 
conventional invisible axion with $F_a\simeq 10^{12}$ GeV 
at low energy in models with
an anomalous $U(1)$ gauge group.  With an extra nonabelian gauge
group, it is more difficult to make the model-independent axion
the conventional invisible axion.  Finally, we commented the
interesting cosmological scenario if the gravitino mass falls in
the eV range, which is the first explicit example of cosmology with
late decaying particles after the COBE data.  

\section*{Acknowledgements}
I thank YITP for the kind hospitality extended to me during my
visit.  This work is supported in part by the Korea Science and
Engineering Foundation through Center for Theoretical Physics,
Seoul National University, Korea-Japan Exchange Program, the Ministry
of Education through the Basic Science Research Institute,
Contract No. BSRI-94-2418, and SNU Daewoo Research Fund.


\begin{thebibliography}{99}
\bibitem{rf:gut} H. Georgi, H. R. Quinn and S. Weinberg,
Phys. Rev. Lett. {\bf 33}, 451 (1974); J. Pati and Abdus Salam,
Phys. Rev. {\bf D10}, 275 (1974); H. Georgi and S. L. Glashow,
Phys. Rev. Lett. {\bf 32}, 438 (1974).

\bibitem{rf:axion} R. D. Peccei and H. R. Quinn, Phys. Rev.
{\bf D16}, 1791 (1977); S. Weinberg, Phys. Rev. Lett. {\bf 40},
223 (1978); F. Wilczek, Phys. Rev. Lett. {\bf 40}, 279 (1978);
J. E. Kim, Phys. Rev. Lett. {\bf 43}, 103 (1979); A. P.
Zhitnitskii, Sov. J. Nucl. Phys. {\bf 31}, 260 (1980); M. A. 
Shifman, V. I. Vainstein and V. I. Zakharov, Nucl. Phys.
{\bf B166}, 4933 (1980); M. Dine, W. Fischler and M. Srednicki,
Phys. Lett. {\bf B104}, 199 (1981).

\bibitem{rf:tech} S. Weinberg, Phys. Rev. {\bf D19}, 1277 (1979);
L. Susskind, Phys. Rev. {\bf D20}, 2619 (1979).

\bibitem{rf:susy} E. Witten, Nucl. Phys. {\bf B188}, 513 (1981).

\bibitem{rf:u1} See, for example, L. Ibanez and G. Ross, Phys.
Lett. {\bf B332}, 110 (1994).

\bibitem{rf:rev} J. E. Kim, Phys. Rep. {\bf 150}, 1 (1987); 
H.Y. Cheng, Phys. Rep. {\bf 158}, 1 (1988); R. D. Peccei, in
{\it CP Violation}, ed. C. Jarlskog (World Scientific, Singapore,
1989).

\bibitem{rf:inst} A. A. Belavin, A. Polyakov, A. Schwartz, and Y. 
Tyupkin, Phys. Lett. {\bf B59}, 85 (1974).  

\bibitem{rf:edm} I. S. Altarev et al, Phys. Lett. {\bf B276}, 242 (1992). 

\bibitem{rf:axcos} J. Preskill, M. B. Wise and F. Wilczek, Phys.
Lett. {\bf B120}, 127 (1983); L. F. Abbott and P. Sikivie,
Phys. Lett. {\bf B120}, 133 (1983); M. Dine and W. Fischler,
Phys. Lett. {\bf B120}, 137 (1983).

\bibitem{rf:uq} D. B. Kaplan and A. Manohar, Phys. Rev. Lett.
{\bf 56}, 2004 (1986); K. Choi, C. W. Kim and W. K. Sze, Phys. Rev.
Lett. {\bf 61}, 794 (1988); K. Choi, Nucl. Phys. {\bf B383}, 58 (1992). 

\bibitem{rf:vafa} C. Vafa and E. Witten, Phys. Rev. Lett. {\bf 53},
535 (1984).

\bibitem{rf:current} W. A. Bardeen and S.-H. H. Tye, Phys. Lett. {\bf 
B74}, 229 (1978); V. Baluni, Phys. Rev. {\bf D19}, 2227 (1979).

\bibitem{rf:staxion} E. Witten, Phys. Lett. {\bf 149}, 351 (1984).

\bibitem{rf:compaxion} J. E. Kim, Phys. Rev. {\bf D31}, 1733 (1985), 
K. Choi and J. E. Kim, Phys. Rev. {\bf D32}, 1828 (1985).

\bibitem{rf:pol} G. D. Coughlan, W. Fischler, E. W. Kolb,
S. Raby and G. G. Ross, Phys. Lett. {\bf B131}, 59 (1983).

\bibitem{rf:mu} J. E. Kim and H. P. Nilles, Phys. Lett. {\bf B138},
150 (1984).

\bibitem{rf:gm} G. F. Giudice and A. Masiero, Phys. Lett. {\bf B206},
480 (1988); I. Antoniadis, E. Gava, K. S. Narain and
T. R. Taylor, Nucl. Phys. {\bf B432}, 187 (1994).

\bibitem{rf:musym} J. E. Kim and H. P. Nilles, Mod. Phys. Lett.
{\bf A9}, 3575 (1994).

\bibitem{rf:choi} K. Choi and J. E. Kim, Phys. Lett. {\bf B154},
393 (1985).

\bibitem{rf:choik} K. Choi and J. E. Kim, Phys. Lett. {\bf B164},
71 (1985).

\bibitem{rf:iq} M. Dine, N. Seiberg and E. Witten, Nucl. Phys.
{\bf B289}, 317 (1987); J. J. Atick, L. J. Dixon and A. Sen,
Nucl. Phys. {\bf B292}, 109 (1987); M. Dine, I. Ichinose and
N. Seiberg, Nucl. Phys. {\bf B293}, 253 (1987);
A. Font, L. E. Ibanez, H. P. Nilles and F. Quevedo, Nucl. Phys.
{\bf B307}, 109 (1988); J. A. Casas, E. K. Katehou and C. Munoz,
Nucl. Phys. {\bf B317}, 171 (1989).

\bibitem{rf:mich} J. E. Kim, Phys. Lett. {\bf B207}, 434 (1988).

\bibitem{rf:chun} E. J. Chun, J. E. Kim and H. P. Nilles, Nucl. Phys.
{\bf B370}, 105 (1992).

\bibitem{rf:eps} J. E. Kim, to be published.

\bibitem{rf:rtw} K. Rajagopal, M. S. Turner and F. Wilczek, Nucl.
Phys. {\bf B358}, 447 (1991).

\bibitem{rf:saxino} J. E. Kim, Phys. Rev. Lett. {\bf 67}, 3465
(1991).

\bibitem{rf:sw} S. Weinberg, Phys. Rev. Lett. {\bf 48}, 1303 (1982).

\bibitem{rf:ekn} J. Ellis, J. E. Kim and D. V. Nanopoulos,
Phys. Lett. {\bf B145}, 181 (1984). For a recent study, see,
M. Kawasaki and T. Moroi, Prog. Theo. Phys. {\bf 93}, 879 (1995);
T. Moroi, Thesis (1995) hep-ph/9503210.

\bibitem{rf:fis} W. Fischler, Phys. Lett. {\bf B332}, 227 (1994);
R. G. Leigh and R. Rattazzi, Phys. Lett. {\bf B352}, 20 (1995);
H. Fujisaki, K. Kumekawa, M. Yamaguchi and M. Yoshimura, 
hep-ph/9511381.

\bibitem{rf:dn} M. Dine and A. E. Nelson, Phys. Rev. {\bf D48},
1277 (1993).

\bibitem{rf:bbe} J. Bardeen, J. R. Bond, and G. Efstathiou, Astro. Phys.
J. {\bf 321}, 28 (1987).

\bibitem{rf:ckk} E. J. Chun, H. B. Kim and J. E. Kim, Phys.
Rev. Lett. {\bf 72}, 1956 (1994); H. B. Kim and J. E. Kim,
Nucl. Phys. {\bf B433}, 421 (1995).

\bibitem{rf:nu} M. S. Turner, Phys. Rev. Lett. {\bf 72}, 3754 (1994); 
R. N. Mohapatra and A. Riotto, Phys. Rev. Lett. {\bf 73}, 1324 (1994);
A. D. Dolgov, S. Pastor and J. W. F. Valle, preprint astro-ph/9506011.

\bibitem{rf:mixed} E. L. Wright et al., Astrophys. J. {\bf 396},
L13 (1992).

\bibitem{rf:davis} M. Davis, G. Efstathiou, C. S. Frenk, and
S. D. M. White, Nature {\bf 356}, 489 (1992). 

\end{thebibliography}
\end{document}